\documentclass[conference]{IEEEtran}
\usepackage[pdftex]{graphicx}
\usepackage[usenames,dvipsnames,table,xcdraw]{xcolor}
\usepackage{array}
\usepackage{multirow}
\usepackage[normalem]{ulem}
\useunder{\uline}{\ul}{}
\usepackage{booktabs}
\usepackage{supertabular,booktabs}
\usepackage{longtable}
\usepackage{lscape}
\usepackage{cite}
\usepackage{url}
\usepackage{hyperref}
\usepackage{dirtytalk}
\usepackage{graphicx}
\usepackage{comment}
\usepackage{makecell}
\usepackage{svg}
\usepackage{amsmath}

\usepackage[caption=false]{subfig}
\graphicspath{{Images/}}
\usepackage[normalem]{ulem}
\useunder{\uline}{\ul}{}

\usepackage{tabularx}
\usepackage{balance}
\usepackage{nameref}
\begin{document}

\title{Testing Is Not Boring: Characterizing Challenge in Software Testing Tasks}

\author{

\IEEEauthorblockN{Davi Gama Hardman}
\IEEEauthorblockA{CESAR\\
 Recife, PE, Brazil \\
 dgh@cesar.org.br}
\and

\IEEEauthorblockN{César França}
\IEEEauthorblockA{CESAR School\\
 Recife, PE, Brazil \\
accf@cesar.school}
\and

\IEEEauthorblockN{Brody Stuart-Verner}
\IEEEauthorblockA{University of Calgary\\
Calgary, AB, Canada \\
brody.stuartverner@ucalgary.ca}
\and

\IEEEauthorblockN{Ronnie de Souza Santos}
\IEEEauthorblockA{University of Calgary\\
Calgary, AB, Canada \\
ronnie.desouzasantos@ucalgary.ca} 

}

%{Shell \MakeLowercase{\textit{et al.}}: Bare Demo of IEEEtran.cls for Computer Society Journals}

\IEEEtitleabstractindextext{%
\begin{abstract}
As software systems continue to grow in complexity, testing has become a fundamental part of ensuring the quality and reliability of software products. Yet, software testing is still often perceived, both in industry and academia, as a repetitive, low-skill activity. This perception fails to recognize the creativity, problem-solving, and adaptability required in testing work. Tasks such as designing complex test cases, automating testing processes, and handling shifting requirements illustrate the challenges testing professionals regularly face. To better understand these experiences, we conducted a study with software testing professionals to explore the nature of challenging tasks in software testing and how they affect these professionals. Our findings show that tasks involving creativity, ongoing learning, and time pressure are often seen as motivating and rewarding. On the other hand, a lack of challenge or overwhelming demands can lead to frustration and disengagement. These findings demonstrate the importance of balancing task complexity to sustain motivation and present software testing as a dynamic and intellectually engaging field.
\end{abstract}

}

\maketitle

\IEEEdisplaynontitleabstractindextext

\IEEEpeerreviewmaketitle

\section{Introduction}
\label{sec:introduction}

The increasing complexity of software features, driven by the need to provide users with a variety of innovative services, is making quality assurance increasingly challenging \cite{hirayama2005analysis}. Software testing, the primary technique employed to verify and validate software functionality, is at the forefront of ensuring software quality \cite{lima2023software}. This process guarantees that the software behaves as expected by clients and users based on the given requirements and specifications, identifying faults to be addressed before delivery. Software testing is inherently complex, involving numerous demanding activities \cite{isha2014software, hourani2019impact} that can increase the challenge of the work, especially within the dynamic context of agile development \cite{florea2018software}.

In general, challenging tasks often require professionals to deviate from their routines and adapt their skills, allowing them the freedom to decide how to approach a given assignment \cite{preenen2011managing}. Such tasks are demanding and stimulating and require determination and a high level of responsibility from professionals. A key feature of these tasks is that they tend to attract individuals who are eager to develop a wide range of capabilities and soft skills to support their work \cite{de2009employees}. Studies have shown that challenging tasks promote intrinsic motivation, enhance leadership skills, improve job engagement, and increase job satisfaction. Additionally, they help professionals achieve greater creativity and innovative performance \cite{derue2009developing, preenen2011managing, uphill2019challenge}.

The relationship between challenging tasks and motivated performance has been widely explored in the literature, with scientific evidence showing that challenges elicit a range of cognitive, physiological, behavioral, and emotional responses in potentially stressful scenarios \cite{hase2019relationship}. In this context, challenges are typically associated with positive emotions and seldom linked to negative ones, indicating that challenging tasks are generally beneficial for professionals \cite{uphill2019challenge, jones2009theory}. However, despite the extensive research on the impact of challenging tasks on professional experiences, there is a notable gap in empirical knowledge regarding the specific characteristics of a challenging task in software development, particularly when considering specific activities such as those related to software testing.

Software testing is frequently misunderstood among industry practitioners as a simple, unchallenging task, often seen as purely technical and repetitive \cite{de2023myths, de2017would, lizama2019unpopularity}. This perception overlooks the intricate human aspects of testing, such as the creativity involved in designing test cases and the problem-solving skills needed to anticipate various scenarios. This misconception is also prevalent in academia, where many students view testing as monotonous and lacking professional growth opportunities \cite{de2023myths}, overlooking the demands of tasks like test automation, which call for both technical expertise and innovative thinking.

Therefore, considering that testing is often misinterpreted as a non-challenging activity among software professionals \cite{de2023myths, lizama2019unpopularity}, this study aims to identify and explore the characteristics of challenging activities in software testing to understand their effects on software testing professionals and other practitioners involved in these tasks. To guide our research and its application in an industrial/practical context, we formulated the following questions:

\smallskip \smallskip
{\narrower \noindent \textit{\textbf{RQ1.} What defines a challenging task in software testing, and how do software testing professionals perceive them?} \par}
\smallskip \smallskip

\smallskip \smallskip
{\narrower \noindent \textit{\textbf{RQ2.} What emotions or feelings do challenging tasks evoke in software testing professionals?} \par}
\smallskip \smallskip

Answering these questions has direct practical applications, helping project managers devise interventions that motivate software testing professionals through challenging tasks and subsequently enhance software quality. From this introduction, this study is organized as follows. In Section 2, we present a literature review on challenging tasks in software engineering. Section 3 describes the method conducted in this study. In Section 4, we present our findings, which are discussed in Section 5, along with the implications and limitations of this study. Finally, Section 6 summarizes our contributions and final considerations.

\section{Background} \label{sec:background}
The concept of challenge has been defined over time with various attributes, including emotion, attitude, and state. As a motivational state, challenge is characterized by the subconscious assessment that personal resources meet or exceed situational demands, while threat arises when resources are perceived as insufficient to meet these demands \cite{uphill2019challenge, blascovich2000challenge}. Scientific evidence shows that challenge and threat elicit different cognitive, physiological, behavioral, and emotional responses in stressful situations \cite{uphill2019challenge, hase2019relationship, blascovich2000challenge}. Some authors argue that these states lead to distinct emotional reactions, with challenges being linked to positive emotions and threats to negative emotions \cite{chadha2019, jones2009theory}, viewing them as opposite ends of a continuum.

Traditionally, challenge and threat have been studied in motivated performance situations [14], where individuals must complete tasks under stress to achieve important outcomes \cite{blascovich2000challenge}. Research suggests these psychological states affect effort, attention, decision-making, physical functioning, and overall performance \cite{jones2009theory}. Challenging tasks, in particular, are associated with intrinsic motivation, leadership development, work engagement, job satisfaction, career advancement, and employee retention. They also foster creativity and innovation, as people are often drawn to challenging activities, seeking to further develop their skills and abilities \cite{baer2003rewarding, derue2009developing}.

Most studies on the challenge activities have focused on athletes, students, and workers, in general, \cite{uphill2019challenge, hase2019relationship}, and a review of human factors in software development identified five studies that examined challenges as a motivational factor \cite{dutra2021human}. However, there is a notable gap in research on challenges and their associated feelings, specifically among software testing engineers. Addressing this gap could provide valuable insights and contribute to interventions that promote more engaging and challenging activities in this field, as well as help demystify the idea that software testing is monotonous and lacks challenges \cite{de2023myths, lizama2019unpopularity}. By exploring the complexities and motivations behind testing, we can reshape perceptions and highlight the creative and problem-solving aspects that make it an intellectually demanding profession.

\section{Method} \label{sec:method}
Based on the lack of studies on challenging tasks in software engineering, in particular, considering the specific context of software testing, we followed the guidelines to perform cross-sectional surveys in software engineering \cite{easterbrook2008selecting, ralph2020empirical} and developed a questionnaire to collect the experience of software testing professionals about how they experience challenges in their daily work. Below, we describe the details of our method.

\subsection{Industrial Context} \label{sec:context}
Software testing is often perceived as a repetitive process that lacks the problem-solving elements found in programming and coding. However, this perception does not reflect the actual experiences of software testing professionals, particularly in agile environments \cite{de2023myths, stray2022exploring}. The motivation for conducting this survey stemmed from the lead author's 12 years of experience in software testing, during which he observed a lack of approaches and tools to help managers and leaders motivate testing professionals. These observations were made at the Recife Center for Advanced Studies and Systems (CESAR), a well-established and mature software company in South America.

Founded in 1996, CESAR specializes in on-demand software development for clients across sectors such as finance, telecommunications, manufacturing, and services. With nearly three decades of experience, CESAR has become an influential private center addressing complex challenges through digital transformation, agile methodologies, user experience, and software engineering. Its educational arm, CESAR School, has trained over 5,000 individuals. Internationally recognized for its contributions to research and innovation—including in areas like Generative AI—CESAR integrates software testing as a critical component of its quality assurance processes. The role of software testing professionals is essential in more than 70 projects carried out by the company.

In this context, the survey aims to address the gap in understanding the motivational aspects and challenges faced by software testing professionals, particularly in high-impact environments like CESAR. By shedding light on these factors, the study contributes to dispelling misconceptions about software testing and demonstrates the complexity and essential problem-solving skills involved in the role.

\subsection{Questionnaire} \label{sec:question}
We created the questionnaire displayed in Table \ref{tab:questionnaire} organized into five sections. The questions were designed to gather information from software testing professionals on their background, the context in which they work, their definition of a challenge in software testing based on their experience, the level of challenge in their testing activities, and the emotions or feelings triggered by challenging tasks.

\begin{table}
  \caption{Questionnaire}
  \label{tab:questionnaire}

\begin{tabularx}{\linewidth}{p{1.5cm} p{6.5cm}}
\toprule
Section & Questions \\ \hline 

I. Consent & I declare that I have read the research purposes and that I am aware of the objective of this questionnaire as well as the confidentiality of the data informed by me in this research. \newline ( ) YES \\

II. Background & 
1. Please, provide us with your: \newline
  a. Age \newline
  b. Gender \newline
  c. Location (country/state/city) \newline
  d. Highest level of education \newline
\\

III. Work & 
2. What type of company do you work for? \newline
3. How many years of experience do you have working in software development? \newline
4. How long have you been working on your current project? \newline
\\

IV. Challenge & 
5. How do you define a challenge at your work? \newline
6. Do you find your job challenging? Why? \newline
  a. If yes: What tasks make your job challenging? Could you give us an example? \newline
  b. If not: What would it take for you to find your job challenging? Could you give an example? \newline
7. What characteristics do these tasks have that make you find them challenging? Could you give an example? \newline
8. Can we say that tasks that do not have all of these characteristics mentioned by you are not challenging tasks? Why? \newline
\\

V. Emotions & 
9. How do you feel when you face a challenging task at work? Could you describe at least one recent example in the work context where this happened? \newline
10. How do you feel when you face an extremely challenging task at work? Could you describe at least one recent example in the work context where this happened? \newline
11. How do you feel about a task that is not challenging? Could you describe at least one recent example in the work context where this happened? \newline
\\

\bottomrule
\end{tabularx}

\end{table}

\subsection{Data Collection} \label{sec:participants}
The study population comprised a diverse group of software testing professionals, including test engineers, test analysts, testing managers, and QA analysts, among others, all of whom are directly involved in testing and quality assurance in software development processes. To ensure a representative sample, we adhered to established sampling recommendations for software engineering research \cite{baltes2022sampling}, starting with convenience sampling to engage participants easily accessible to the researchers. This was supplemented by snowball sampling, where initial participants referred additional professionals within their network.

Given the qualitative nature of this survey, we employed the saturation sampling technique \cite{fontanella2011amostragem} to define the final sample size. Saturation was considered reached when additional participant responses no longer introduced new themes, codes, or perspectives relevant to our research questions. This is an accepted approach for ensuring data sufficiency in exploratory studies, particularly when the focus is on depth rather than generalizability. In our case, we observed thematic saturation after approximately 20 participants, and by the 29th response, no new codes or concepts were emerging from the data. Thus, the sample size was considered sufficient to provide an understanding of the phenomenon under investigation.

Participants were recruited through multiple channels, including email, Slack, LinkedIn, and WhatsApp, to reach a broad and diverse group of professionals. In this process, initially, potential participants were informed about the study’s goals, scope, and methodological approach. Once they expressed interest, they were provided with an online version of the Free and Informed Consent Form (ICF) to ensure voluntary participation. After consenting, participants received a link to the online questionnaire, which collected detailed data regarding their experiences, perspectives on challenges, and other relevant aspects of their roles in software testing. This multi-stage recruitment and consent process ensured both the ethical conduct of the study and the quality of the data collected.

\subsection{Data Analysis} 
\label{sec:analysis}
For data analysis, we first employed descriptive statistics ~\cite{george2018descriptive} to summarize and explore the demographic information collected in the background and work context sections of the questionnaire. This allowed us to understand key participant characteristics, such as age, gender, education level, years of experience, and type of company. These demographic insights provided essential context for interpreting the qualitative data and allowed us to explore potential trends related to experience or organizational type.

Following the quantitative analysis, we conducted a thematic analysis ~\cite{cruzes2011recommended} to examine the rich qualitative data gathered from the participants’ open-ended responses. Thematic analysis is a well-established qualitative approach that helps researchers systematically identify, organize, and interpret patterns or themes within qualitative data. In this study, the process began with open coding, where we carefully read through the responses to generate initial codes that described different aspects of the data. These codes were then refined during the axial coding stage, where we looked for relationships between the codes, grouping them into broader categories that reflected recurring themes. Finally, selective coding was applied to identify the most significant themes that directly answered our research questions, such as how participants defined challenges in software testing, the specific characteristics of challenging tasks, and how these challenges affected their professional experience and emotional responses.

This multi-step coding process enabled us to turn complex qualitative data into actionable insights, ensuring that the findings were grounded in the experiences of the participants. By using both descriptive statistics and thematic analysis, we combined quantitative and qualitative methods to provide a comprehensive understanding of the demographics and the nuanced experiences of software testing professionals. This approach ensured that our analysis not only captured the prevalence of certain characteristics or experiences but also explored the meanings and patterns within the data.

\subsection{Ethics}
All procedures in this study followed institutional ethical guidelines from the first author's university for research involving human participants. Participants were informed about the study objectives, the voluntary nature of their participation, and the confidentiality of their responses prior to completing the questionnaire. While the study does not ask directly about projects and technologies being developed by the participants, the final dataset includes personal reflections in a professional setting, where participants often referred to specific projects, clients, and colleagues. To protect their privacy and prevent potential identification, the full dataset cannot be publicly shared. However, we include carefully selected anonymized quotations in the results section to support the transparency and credibility of our analysis.

\section{Findings} 
\label{sec:findings}
Our sample consisted of 29 software testing professionals, ranging in age from 20 to 47 years. While all participants were based in Brazil, many had experience working with international clients, which contributed to a diverse set of experiences in software engineering methodologies, technology domains, and testing contexts, mitigating some of the potential cultural limitations of having a sample from a single country. Participants' experience in software development varied widely, spanning from 1 to 17 years, with an average of 7.6 years and a standard deviation of 5.3 years, indicating a broad spectrum of professional backgrounds. We had 8 participants who identified as women, adding gender diversity to our sample and enriching the range of perspectives.

\subsection{Characterization of Challenge from the Perspective of Testing Professionals}

Before examining the characteristics of tasks that are specifically challenging in software testing, we first explored how testing professionals themselves conceptualize the notion of challenge more broadly. This initial step provided a foundation for understanding how challenges are perceived within their day-to-day work. According to participants, challenges in software development are generally associated with three main features: novelty, the need for adaptation, and increased knowledge demands. These characteristics consistently emerged across accounts, regardless of the specific role, activity, or domain.

The first core element, \textit{novelty}, refers to tasks that introduce something new or unfamiliar into the professional’s routine. Participants described such tasks as especially demanding because they disrupt established workflows and require creative thinking to navigate. Rather than relying on repetitive patterns, professionals must engage with unfamiliar problems and devise new approaches, which adds complexity and makes the task more stimulating.

The second element, \textit{adaptation}, reflects the need to adjust quickly to new or evolving circumstances. Participants emphasized that challenging tasks often push them beyond their usual boundaries, requiring them to leave their comfort zones. Whether adapting to changes in time constraints, technical requirements, or team dynamics, professionals identified flexibility as a key factor in managing and overcoming these challenges.

The third defining feature is the \textit{knowledge demand} involved in difficult tasks. Participants frequently noted that tasks which compel them to acquire new knowledge or skills are inherently more challenging, but also more rewarding. These tasks stretch their existing capabilities and promote professional growth, particularly when the demands extend beyond their current knowledge base.

Together, these three characteristics, novelty, adaptation, and knowledge demand, shape how testing professionals perceive challenges in software development. Table \ref{tab:charac} summarizes these findings. This general understanding provides important context for examining the specific challenges encountered in the practice of software testing.

\begin{table}
  \caption{Characteristics of Challenge}
  \label{tab:charac}

\begin{tabularx}{\linewidth}{p{1.5cm} p{2.3cm} X}
\toprule
Characteristic & Definition & Evidence Examples \\ \hline 

Novelty & Challenge is characterized by introducing new tasks into the routine, typically requiring creativity to complete. & P2: \textit{``The new or the unknown [. . . ] what comes out of the
monotonous.''} \newline P4: \textit{``something that is different from your context and that requires more time for study and planning.''}  \newline P29: \textit{``It is something new, a way to reach the final goal that requires effort, creativity, and skill.''} \newline \\

Adaptation & Challenge is characterized by requiring transformation and change in professionals' routines. & P5: \textit{``Do something that goes beyond my limits and comfort zone.''} \newline P7: \textit{``[. . . ] it is what requires easy adaptation and quick learning.''}  \newline P13: \textit{``I challenge myself and everything that makes me leave my comfort zone.''} \newline \\

Knowledge \newline Demanding & Challenge is characterized by demanding learning, enabling professionals to improve their capabilities. & P20: \textit{``things that require knowledge that you do not yet have.''} \newline P21: \textit{``Everything that goes beyond my knowledge.''}  \newline P28: \textit{``Activities that I have not yet fully mastered.''} \\

\bottomrule
\end{tabularx}

\end{table}

\subsection{Challenge in Software Testing Activities}

In the context of software testing, professionals identified several specific factors that contribute to the challenges they face, which can be grouped into four main dimensions: skill, environment, attitude, and task. Each dimension encompasses multiple elements that increase the complexity of the work and require testers to constantly adapt to fulfill their responsibilities.

\textbf{Skill Dimension}. The \textit{skill} dimension captures the technical and cognitive demands placed on testing professionals. A key challenge in this area is \textit{knowledge acquisition}. Participants emphasized the need to constantly learn new tools, technologies, and testing techniques to keep pace with industry developments. While this ongoing learning is essential for designing effective testing strategies, it also imposes a cognitive load, as professionals must frequently pause their work to study, investigate, and plan. Another important factor is \textit{communication}. Testers must interpret a range of technical artifacts, including requirements documents, design specifications, and user stories, and clearly communicate issues and improvements with developers, project managers, and clients. Miscommunication at any stage can lead to misunderstandings that compromise the testing process. A third aspect of this dimension is \textit{knowledge dissemination}. Participants noted that they are not only responsible for acquiring knowledge but also for sharing it with colleagues, especially in collaborative environments. This requires testers to explain complex concepts clearly and accommodate different learning styles within the team, which can be demanding and time-consuming.

\textbf{Environment Dimension}. The \textit{environment} dimension refers to external conditions that shape the testing process. One of the most frequently cited challenges in this category is \textit{instability}. Participants described frequent changes in project requirements, shifting priorities, and last-minute decisions that forced them to revise or redesign testing strategies, often when development was already at an advanced stage. This unpredictability disrupted their ability to maintain a consistent workflow. Another prominent challenge was \textit{time constraints}. Delays in previous development stages often reduced the time available for thorough testing, putting testers under pressure to meet deadlines without compromising quality. Finally, the \textit{lack of resources}, such as missing or outdated documentation, added to the difficulty. Testers described having to work with incomplete information, making it harder to define test strategies and identify meaningful scenarios, which increased the risk of overlooking critical defects.

\textbf{Attitude Dimension}. The \textit{attitude} dimension involves personal and behavioral factors that influence the testing process. One of the main challenges in this area is \textit{individual responsibility}. Participants noted that they are expected to take ownership of the whole quality process, often bearing significant accountability for the final outcome of the product. This expectation demands a high degree of autonomy and judgment. Alongside this, \textit{individual commitment} was consistently mentioned. Testers must remain highly engaged and motivated, particularly when facing complex tasks or compressed timelines. The level of dedication required involves active collaboration with other team members, continuous self-monitoring of testing practices, and maintaining a strong sense of ownership over the work, even under difficult conditions.

\textbf{Task Dimension}. The \textit{task} dimension centers on the specific activities carried out during software testing. One notable challenge here is \textit{task interchangeability}. Participants described the frequent need to switch between different types of tasks, such as writing test cases, executing test plans, analyzing defects, and reporting issues, which requires constant shifts in focus and context. This can lead to cognitive overload, as testers must sustain high levels of attention while managing diverse responsibilities. Another significant challenge is \textit{task complexity}. Many testing tasks involve interacting with complex system features or scenarios that lack predefined solutions. In these cases, testers must rely on critical thinking and creativity, since simple lookups or scripted answers are insufficient. These complex tasks demand a deep understanding of both the system under test and the broader testing strategy, increasing the cognitive and technical demands placed on professionals.

In general, our findings demonstrate that challenges in software testing emerge from an interplay of technical, environmental, personal, and task-related factors. The constant need to learn and share knowledge, adjust to changing conditions, manage high-pressure timelines, and navigate cognitively demanding activities makes software testing a dynamic and intellectually engaging profession. These dimensions, summarized in Table \ref{tab:challenge}, illustrate the multifaceted nature of the testing process and emphasize the importance of a skilled, adaptable, and committed workforce.

\begin{table*}
  \caption{Challenging Testing Tasks}
  \label{tab:challenge}

\begin{tabularx}{\linewidth}{p{1.5cm} p{2cm} p{5cm} X}
\toprule
Dimension & Factor & Definition & Evidence Examples \\ \hline 

Skill & Knowledge Acquisition & Software testing professionals must continuously learn and experiment to stay abreast of technologies, techniques, and tools, ensuring the process remains up to date. & P02: \textit{``[...] always having new things to learn.''} \newline P04: \textit{``You need to study and stop to plan and understand the context.''}  \newline P06: \textit{``[...] definition of testing strategies, which involves concepts, knowledge of technologies, processes, and tools.''} \newline \\

 & Communication & Software testing professionals must interpret various artifacts and effectively communicate issues with both the team and the client. & P08: \textit{``Keep the team informed regarding improvements in the test development process.''} \newline P14: \textit{``[...] have good communication skills.''}  \newline P22: \textit{``[...] report to the customer.''} \newline \\

& Knowledge \newline Dissemination & Software testing professionals must actively share knowledge with their peers and support collaboration across quality activities. & P02: \textit{``Guiding people is a challenge, [...] because people learn in different ways.''} \newline \\ \hline 

Environment & Instability & Uncertainties of requirements, priorities, or decisions can cause significant changes in the testing strategy, particularly when they arise in the advanced stages of the development cycle. & P07: \textit{``Changes that occurred when development was at an advanced stage or already completed.''} \newline P25: \textit{``I’m always working with different software versions, and new problems always happen.''}  \newline P26: \textit{``[...] new challenges are presented to us, new priorities, new types of processes and methodologies, new work tools and new ways to solve problems of the most varied types.''} \newline \\

 & Time & Delays in earlier stages of software development, such as programming, often lead to a significant reduction in the time allocated for testing activities. & P06: \textit{``Deadlines often become a relevant challenge in the quality process, and convincing the customer is not always an easy task.''} \newline P10: \textit{``Deadline problems always keep the challenge.''}  \newline P27: \textit{``[...] do everything within the deadline''} \newline \\

& Resources & The absence or lack of resources, such as documentation artifacts, complicates the process of defining strategies, identifying test scenarios, and creating test cases. & P06: \textit{``understand the business rules and technologies involved, most of the time, with little or no documentation.''} \newline P18: \textit{``[lack] of documentation.''} \newline \\ \hline 

Attitude & Individual \newline Responsibility & Pro-activeness to ensure the process remains operational and efficient in terms of quality. & P05: \textit{``[...] high degree of responsibility and trust given to me by the team.''} \newline P24: \textit{``High cost of development and responsibility for the results.''}  \newline P30: \textit{``They require creativity and technical knowledge for the result to be satisfactory.''} \newline \\

 & Individual \newline Commitment & The importance of software testing professionals collaborating with their team.. & P07: \textit{``This requires a high degree of commitment and agility from the QA team.''} \newline P21: \textit{``New features always require [...] involvement so that the tests are well-designed and executed.''} \newline \\ \hline 

Task & Task \newline Interchangeability & The frequency of professionals switching activities increases the challenge, as changing contexts too often demands high levels of concentration. \newline & P05: \textit{``[...] I usually don’t spend two days in a row doing the same activity.''} \newline \\

 & Task \newline Complexity & The importance of software testing professionals collaborating with their team. & P23: \textit{``These are not tasks with such direct answers or solutions, so it’s not just a matter of searching on Google or StackOverflow, for example.''} \newline P29: \textit{``Complex system features [...].''} \\

\bottomrule
\end{tabularx}

\end{table*}

\subsection{The Effects of Challenging Tasks on Software Testing Professionals}

Our findings indicate that workplace challenges are multifaceted and have a significant impact on software testing professionals. These challenges evoke a wide range of emotional responses, which play a critical role in shaping motivation and job satisfaction. Understanding this emotional landscape is essential for leaders and managers, as unmanaged or poorly distributed challenges may lead skilled professionals to disengage or leave their roles. Moreover, recognizing these dynamics helps challenge outdated perceptions of software testing as monotonous or low-skill work.

Participants described several positive effects that challenging tasks can produce. These include feelings of excitement, confidence, curiosity, determination, engagement, and happiness. Many professionals reported that challenges increase their motivation, making them feel more involved and eager to contribute to their projects. The opportunity to witness personal growth and skill development in response to these tasks further reinforces positive emotions, making challenging work a source of professional fulfillment.

However, participants also acknowledged the potential for negative emotional responses. Reported effects included anxiety, apprehension, disinterest, stress, and worry. For some, the fear of not meeting expectations or the uncertainty associated with complex tasks introduced significant stress. If not addressed, these negative emotions can diminish motivation and, over time, contribute to burnout.

These findings underscore the importance of managerial awareness in shaping how challenges are introduced and managed. It is crucial to strike a balance—offering enough complexity to stimulate motivation and growth, without overwhelming individuals or compromising well-being. By recognizing how different professionals respond to challenges, managers can better tailor task difficulty to promote engagement and performance while minimizing adverse effects. Table \ref{tab:effects} summarizes these findings, categorizing the effects of challenges into positive and negative emotional responses.

\begin{table*}
  \caption{Effects of Challenge on Testing Professionals}
  \label{tab:effects}
\scriptsize

\begin{tabularx}{\linewidth}{p{1.5cm} p{5cm} X}
\toprule
Effect & Emotions & Evidence Examples \\ \hline

Positive & Challenges have the potential to elevate excitement, confidence, curiosity, determination, disposition, and happiness. & P6: \textit{`` the challenges usually motivate me more than demotivate.''} \newline P10: \textit{``Motivated, especially when it is possible to see the evolution and growth in the face of this challenge.''}  \newline P25: \textit{``I feel excited and willing to contribute more and more to
the project.''} \newline \\

Negative & Challenges may induce feelings of anxiety, apprehension, disinterest, stress, and worry. & P09: \textit{``[. . . ] worried about the results not meeting the desired level.''} \newline P13: \textit{``I am VERY afraid of not being able to perform well.''}  \newline P28: \textit{``Stressed and anxious, due to the doubt and uncertainties.''} \\

\bottomrule
\end{tabularx}

\end{table*}

\section{Discussion} \label{sec:discuss}

The experiences shared by participants align with existing definitions of workplace challenges found in the literature \cite{preenen2011managing, weissinger1995development, kirby2014challenge}. In the broader context of the software industry, challenges are typically associated with tasks that involve novelty, complexity, and the need for acquiring or refining knowledge and skills. Contrary to outdated perceptions of software testing as monotonous or repetitive, our findings show that testing involves a wide range of challenging tasks that demand creativity, problem-solving, and critical thinking, skills often associated with programming and planning activities \cite{de2023myths, lima2023software}. Tasks such as test automation, dealing with complex system features, and managing incomplete requirements require testers to think critically and adapt to dynamic project conditions, directly challenging the misconception that testing lacks intellectual engagement.

Our study also reinforces long-standing concerns within the field. A significant issue raised by participants is the imbalance in time allocation between programming and testing, a well-documented source of tension among testers, developers, and project managers \cite{cohen2004managing}. Testing professionals often face compressed timelines due to delays earlier in the development cycle, which heightens anxiety, stress, and apprehension. This is consistent with prior findings that testers experience more time pressure than developers \cite{shah2014global}. These results underscore the importance of project management practices that recognize the complexity of testing and ensure adequate time is allocated for it.

Another prominent challenge identified is the persistent lack of resources to support testing activities, which often contributes to project instability. This aligns with previous studies that have linked resource constraints and shifting requirements—driven by evolving market demands or organizational changes—to increased testing difficulty \cite{nurmuliani2004analysis}. While participants acknowledged that such conditions can foster learning and growth, they also warned that excessive pressure and unpredictability may lead to burnout. This highlights the need for balanced management strategies that offer both challenge and support.

In addition, our findings reveal a strong connection between interpersonal (soft) skills and the challenges encountered in testing. This reflects industry trends that increasingly value testers who possess strong communication, analytical thinking, conflict resolution abilities, and adaptability \cite{florea2018software}. Participants reported that engaging with challenging tasks often helped them develop these competencies, suggesting that such experiences are not only technically enriching but also professionally transformative. Success in testing thus requires not only technical expertise but also the ability to navigate team dynamics and communicate effectively.

Broadly, our findings challenge the traditional view of software testing as a routine or low-skill activity. Instead, they portray testing as a demanding and intellectually stimulating aspect of software development. Like programming, testing requires professionals to navigate complexity, adapt to unstable environments, and apply creative problem-solving, making it a dynamic part of modern software engineering practice.

\subsection{Implications and Lessons Learned } \label{sec:implications}
This study provides important implications for industry practice in software testing, particularly in how challenges are characterized, perceived, and managed within software teams. Drawing from our findings and their alignment with prior literature, we reaffirm that software testing is far from a repetitive or low-skill activity. Testing professionals routinely engage in complex, intellectually demanding tasks that require creativity, critical thinking, and adaptability—especially when automating tests, improving test coverage, working with incomplete or volatile requirements, and managing tight deadlines \cite{de2023myths, lima2023software, cohen2004managing}.

These findings call for a shift in how the industry and educational institutions perceive and support software testing roles. Leaders and educators should work to challenge outdated views and recognize the essential contributions that testing professionals make to the software development process. Based on our results and the broader literature, we offer the following recommendations for practitioners:

\begin{itemize}
\item Rotate testing professionals across a variety of tasks regularly to provide diverse learning experiences and reduce task monotony, fostering engagement and skill development.
%\item Schedule regular one-on-one meetings with testing professionals to proactively address demotivating factors—both personal and team-related—and support well-being and performance.
\item Consistently highlight the value of testing professionals’ contributions to project success, client satisfaction, and societal outcomes to reinforce their sense of purpose and professional identity.
\item Encourage continuous learning by supporting access to the latest technologies and methodologies, helping foster a culture of innovation and ongoing improvement.
\item Invest in training programs that strengthen both technical and interpersonal skills, with an emphasis on communication, knowledge-sharing, and collaboration.
\item Promote the use of automation to reduce repetitive tasks, enabling testers to focus on strategic and intellectually stimulating responsibilities.
\item Set clear, measurable goals and provide timely recognition and rewards to maintain motivation and acknowledge achievements.
\end{itemize}

We anticipate that adopting these recommendations will enhance the job satisfaction of testing professionals while contributing to a broader cultural shift in how the testing profession is viewed. Elevating testing as a dynamic and intellectually engaging component of software engineering can help attract, retain, and motivate skilled professionals and reinforce the value of their work within the software development lifecycle.

\subsection{Threats to Validity} \label{sec:limitations}
While our study offers meaningful results, some limitations should be acknowledged. First, with a sample size of 29 participants and a qualitative approach, we do not claim statistical generalizability. However, this sample size aligns with accepted practices in qualitative research, such as grounded theory and case study methods commonly used in software engineering. Data collection was concluded when thematic saturation was reached, that is, when no new themes emerged. Our goal is not generalization, but transferability: the ability for findings to be re-analyzed or applied in similar contexts. Second, all participants were based in Brazil, which may constrain the diversity of cultural perspectives. To mitigate this, we selected professionals who work with international clients and follow globally adopted software processes, offering broader insights beyond local practices. Finally, as with all qualitative research, there is a risk of researcher bias during analysis. We addressed this through iterative coding and collaborative discussions among researchers to ensure consistency and credibility in theme development.

\section{Conclusions}  \label{sec:conclusions}
This study explored how software testing professionals experience challenging tasks in their daily work, identifying three defining characteristics of challenge in this context: novelty, adaptation, and the need for ongoing knowledge acquisition. These challenges are not solely technical, they also influence how professionals perceive their roles, often evoking a mix of excitement, satisfaction, anxiety, and stress. While some challenges can lead to frustration or burnout, many participants described them as opportunities for growth and deeper engagement, especially when supported by leadership practices that balance complexity, recognize contributions, and foster continuous learning. By surfacing these experiences, our study helps reframe software testing as a creative and intellectually demanding activity, rather than a repetitive or low-skill task. These findings demonstrate the importance of treating testing as a key area of challenge, innovation and professional development within software teams. For future work, we plan to broaden the scope of our study to include testing professionals from diverse regions and organizational settings, and to develop frameworks that link the nature of challenging testing tasks with job satisfaction, performance, and long-term retention, advancing our understanding of how testing roles can be better supported as software systems and expectations evolve.

\ifCLASSOPTIONcaptionsoff
  \newpage
\fi

\balance
\bibliographystyle{IEEEtran}
\bibliography{bib.bib}

\end{document}